\documentclass[aps,prd,onecolumn,12pt,nofootinbib,superscriptaddress,longbibliography,floatfix]{revtex4-2}

\usepackage{amsmath}
\usepackage{amssymb}
\usepackage{dcolumn}
\usepackage{bm}
\usepackage{graphicx}
\usepackage{float}
\usepackage{slashed}
\usepackage{setspace}
\usepackage{xcolor}
\usepackage[colorlinks=true,linkcolor=blue,citecolor=blue,urlcolor=blue]{hyperref}

\newcommand{\Jpsi}{J/\psi}
\newcommand{\psip}{\psi(2S)}

\newcommand{\Upsone}{\Upsilon(1S)}
\newcommand{\Upstwo}{\Upsilon(2S)}

\newcommand{\etac}{\eta_{c}}
\newcommand{\etab}{\eta_{b}}
\newcommand{\etap}{\eta^{\prime}}
\newcommand{\etapp}{\eta^{(\prime)}}

\newcommand{\eps}{\varepsilon}

\begin{document}

\title{\texorpdfstring{Relativistic corrections to $\Upsilon(1S,2S)\rightarrow\gamma\etapp$
in perturbative QCD}{Relativistic corrections to Upsilon(1S,2S) to gamma eta(') in perturbative QCD}}

\author{Cong Wang}
\affiliation{College of Physics and Electronic Science, Hubei Normal University,
Huangshi 435002, China}

\author{Chao-Jie Fan}
\email{fancj@hbnu.edu.cn}
\affiliation{College of Physics and Electronic Science, Hubei Normal University,
Huangshi 435002, China}

\author{Jun-Kang He}
\email{hejk@hbnu.edu.cn}
\affiliation{College of Physics and Electronic Science, Hubei Normal University,
Huangshi 435002, China}

\begin{abstract}
We calculate the radiative decays $\Upsone,\Upstwo\rightarrow\gamma\etapp$ in perturbative
quantum chromodynamics through order $q^{2}$ in the relativistic expansion, including
the quark-content, two-gluon, and electromagnetic contributions. The quark contribution
dominates, with substantial constructive electromagnetic interference and a strongly
suppressed gluon contribution. The annihilation rates depend weakly on the tested
light-cone distribution amplitudes and light-quark masses. Relativistic corrections
reduce the $\Upsone$ rates by a few percent but enhance the $\Upstwo$ rates by
approximately $17\%$--$30\%$. The ratio of the $\gamma\etap$ and $\gamma\eta$ branching
fractions is sensitive to the light-meson mixing parameters, with predictions differing
by about a factor of $5$ between two representative schemes. At a scale of half the
quarkonium mass, however, the $\Upsone\rightarrow\gamma\etap$ central predictions in the two
mixing schemes exceed the experimental upper limit by factors of approximately $5$ and $7$. We also examine
the heavy-pseudoscalar admixtures in the light mesons. The $\etac$ contribution is small,
whereas $\etab$ mixing can substantially modify the rates through interference.
Destructive interference alleviates the discrepancy. With the scale set to the quarkonium
mass, the prediction in the mixing scheme inferred from transition form factors is
compatible with the upper limit within the input uncertainty.
\end{abstract}

\maketitle

%==================================================================
\section{Introduction}
\label{sec:intro}

The radiative decays $\Upsone,\Upstwo\rightarrow\gamma\etapp$ provide a setting in which
to study heavy-quark annihilation into light pseudoscalar mesons. Their amplitudes connect
the dynamics of the initial heavy bound state with the quark and gluon components of the
$\etapp$ and their mixing. Compared with charmonium, bottomonium has a larger heavy-quark
mass and a smaller strong coupling, which motivate a perturbative treatment of the
annihilation process~\cite{QuarkoniumWorkingGroup:2004kpm,Brambilla:2010cs,Voloshin:2007dx}.
These differences also affect the relative importance of the strong and electromagnetic
contributions. The two light-meson channels involve different combinations of flavor
components, so their comparison probes the mixing parameters as well as the underlying
annihilation dynamics. Studying both radial states therefore allows us to examine how the
annihilation picture developed for $\Jpsi,\psip\rightarrow\gamma\etapp$ changes at the
bottomonium scale and how the bound-state corrections affect its predictions.

The experimental situation places a stringent constraint on this picture. In contrast
to the observed charmonium transitions~\cite{ParticleDataGroup:2024cfk}, the CLEO search
in $21.2\times10^{6}$ $\Upsone$ decays found no signal for either channel and obtained
the upper limits~\cite{CLEO:2007iso}
\begin{equation}
\label{eq:cleo}
\mathcal{B}(\Upsone\rightarrow\gamma\eta)<1.0\times10^{-6},\qquad
\mathcal{B}(\Upsone\rightarrow\gamma\etap)<1.9\times10^{-6}\quad(90\%~\mathrm{C.L.}).
\end{equation}
These bounds require branching fractions far below those of the corresponding $\Jpsi$
decays and lie below several theoretical estimates. They raise the question of whether
a perturbative annihilation calculation can account for the suppression once the
different contributions and the relativistic corrections are treated consistently.
The comparison also requires assessing which theoretical inputs and additional
amplitudes can materially affect the predicted rates.

Existing descriptions differ in their treatment of the transition from heavy-quark
annihilation to the light pseudoscalar meson. Quark-model, anomaly-based,
vector-meson-dominance, and factorization approaches give substantially different
estimates~\cite{Korner:1982vg,Kuhn:1983yr,Novikov:1979uy,Intemann:1983yj,Ma:2002ww,Li:2007dq}.
Anomaly-based descriptions relate the light-meson production to gluonic matrix elements,
whereas light-cone factorization resolves the momentum distribution of its constituents.
These treatments can lead to different heavy-quark mass dependences, making the
extrapolation from charmonium sensitive to the underlying assumptions. In the
factorization analysis of Ma~\cite{Ma:2002ww}, the ratio of the bottomonium and charmonium
$\gamma\etap$ branching fractions contains a factor $m_c^6/m_b^6$, illustrating how
strongly the predicted suppression can depend on the heavy-quark masses.

This sensitivity also connects the comparison of the two systems to their bound-state
dynamics. Ma~\cite{Ma:2002ww} identified the relativistic corrections in the $\Jpsi$ decay and the charm-quark mass
as major uncertainties in relating charmonium and bottomonium rates. Using quarkonium
masses was proposed as a phenomenological way to reduce these uncertainties, while
a consistent treatment of the corrections remained necessary. This motivates an
explicit calculation of the relativistic terms in the bottomonium amplitudes rather
than an inference based only on scaling the charmonium rates.

Our earlier perturbative study of $\Jpsi\rightarrow\gamma\etapp$ provides a starting
point for such a calculation. The summed loop amplitudes were found to be infrared
finite and weakly dependent on the light-quark masses and the shapes of the light-cone
distribution amplitudes (DAs)~\cite{He:2019mpy}. The annihilation calculation reproduced
the measured ratio of the two branching fractions for suitable mixing parameters,
while the individual rates remained below the data. This distinction motivates
examining the absolute rates and their ratio together when testing the same framework
in bottomonium. We subsequently included the
order-$q^{2}$ bound-state corrections in the radiative decays of both $\Jpsi$ and
$\psip$~\cite{He:2026yex}. Their substantial effects, particularly for the radially
excited state, showed that the nonrelativistic approximation alone does not establish
the accuracy of the predicted rates. These studies provide a common framework for
comparing the charmonium and bottomonium transitions while retaining the distinct
bound-state inputs and kinematics of each system.

The smaller characteristic heavy-quark velocity in bottomonium suggests a more
favorable relativistic expansion, but the size and sign of a decay correction also
depend on its hard coefficients and interference with the leading amplitude. The
relativistic NRQCD matrix elements extracted from bottomonium leptonic decays indicate
substantially different bound-state effects for the $1S$ and $2S$ states~\cite{Chung:2010vz}, providing a further
reason to examine them separately. A comparison of the two radial states thus probes
the bound-state dependence while retaining the same description of the light mesons.
An explicit calculation can determine whether
the corrections remain subleading in these channels and whether they appreciably
change the comparison with the experimental bounds. At the same time, the different
strong coupling and flavor weights can alter the balance between strong and
electromagnetic amplitudes, so the bottomonium rates need not follow a simple
rescaling of the charmonium results.

To address these questions, we calculate $\Upsone,\Upstwo\rightarrow\gamma\etapp$
in the Bethe-Salpeter framework, expanding the annihilation amplitudes through
order $q^{2}$ and including the quark, two-gluon, and electromagnetic contributions.
The light-meson components are described by their light-cone DAs. Comparing the
individual amplitudes and their interference establishes which contributions control
the rates and how the relativistic terms modify them. We examine the dependence on
the DA inputs and use two representative $\eta$--$\etap$ mixing schemes to distinguish
this dependence from the sensitivity to the mixing parameters. Alongside the
branching fractions, evaluated consistently through order $q^{2}$, we study the ratio
\begin{equation}
\mathcal{R}_{nS}=\frac{\mathcal{B}(\Upsilon(nS)\rightarrow\gamma\etap)}
{\mathcal{B}(\Upsilon(nS)\rightarrow\gamma\eta)},\qquad n=1,2 .
\label{eq:Rdef}
\end{equation}
The common wave-function factor cancels in this ratio, providing a complementary
comparison of the two light-meson channels. Examining the absolute rates and their ratio
together helps distinguish changes shared by both channels from those associated with
their different flavor composition.

We compare the $\Upsone$ predictions with the CLEO bounds and calculate the
corresponding $\Upstwo$ channels in the same framework. To assess effects beyond
the light-quark, gluon, and electromagnetic terms, we also include the $\etac$ and
$\etab$ admixtures of the light mesons and examine their interference with the
annihilation amplitudes. The constructive and destructive choices for the $\etab$
contribution illustrate the dependence on its undetermined relative phase. Together
with a variation of the renormalization scale, this supplementary analysis tests how
these effects can modify the comparison with experiment.

The remainder of this paper is organized as follows. In Sec.~\ref{sec:framework} we set up
the calculation and derive the quark, gluon, and quantum electrodynamics (QED) contributions, including the
relativistic corrections. The numerical analysis is presented in Sec.~\ref{sec:numerical},
and we summarize in Sec.~\ref{sec:summary}.

%==================================================================
\section{Theoretical framework}
\label{sec:framework}

The framework is the one we developed for the charmonium decays
$\Jpsi,\psip\rightarrow\gamma\etapp$~\cite{He:2019mpy,He:2026yex}, carried over to
bottomonium by replacing the annihilating $c\bar c$ pair with a $b\bar b$ pair. The two heavy
sectors differ only in the heavy-quark charge and mass and in the bound-state wave function, so
that the Lorentz and color structures, the loop kernels and their numerators, and the
light-hadron matrix elements transfer unchanged. We therefore set out the ingredients needed to
make the bottomonium calculation self-contained, compress the derivations by referring to
Ref.~\cite{He:2026yex} where they are given in full, and dwell instead on what is specific
to the bottomonium kinematics.

We work in the rest frame of the decaying bottomonium $\Upsilon_{n}$, with $n=1,2$ labeling
$\Upsone$ and $\Upstwo$. Its four-momentum and mass are $K$ and $M$, the photon carries $k$, and
the outgoing pseudoscalar carries $p=K-k$ with mass $m\equiv m_{\etapp}$, so that
$k\!\cdot\!K=(M^{2}-m^{2})/2$. Three short-distance contributions arise, the quark-antiquark,
two-gluon, and QED ones, and we treat them in turn. The treatment of the initial bound state,
common to all three, is set out in Sec.~\ref{subsec:quark} where it first enters. Throughout the
hard kernel we adopt the weak-binding approximation $m_{b}=M/2$. Two ratios recur below, the
squared meson-to-bottomonium mass ratio $x=m^{2}/M^{2}$, which takes the values
$3.4\times10^{-3}$ for $\gamma\eta$ and $1.0\times10^{-2}$ for $\gamma\etap$, an order of
magnitude below the charmonium ones, and $m_{q}^{2}/M^{2}$ with $m_{q}$ the light-quark mass in
the loop.

\subsection{Quark-content contribution}
\label{subsec:quark}

\begin{figure}[H]
\centering
\includegraphics[width=0.5\textwidth]{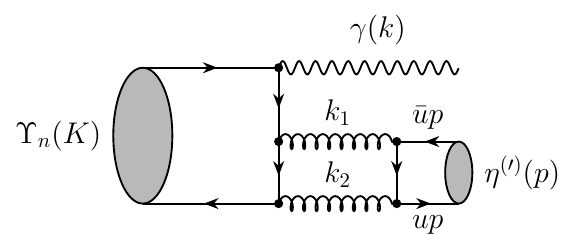}
\caption{A representative Feynman diagram for the quark-content contribution to
$\Upsilon_{n}\rightarrow\gamma\etapp$, in which the $b\bar b$ pair annihilates into a photon
and two virtual gluons that couple to the $q\bar q$ content of the $\etapp$. Five further
diagrams are obtained by permuting the photon and gluon legs.}
\label{fig:quark}
\end{figure}

For the quark content of the $\etapp$ the decay proceeds through the annihilation
$\Upsilon_{n}\rightarrow\gamma g^{*}g^{*}$, the two virtual gluons converting into the $q\bar q$
content of the meson, as in the representative diagram of Fig.~\ref{fig:quark}.

The initial bound state is treated in the Bethe-Salpeter formalism. With the $b$ and $\bar b$
carrying $K/2\pm q$, the $\Upsilon_{n}\rightarrow\gamma g^{*}g^{*}$ amplitude is the
four-dimensional convolution of the Bethe-Salpeter wave function $\Psi(K,q)$ with the hard
kernel $\mathcal{O}(q)$,
\begin{equation}
A=\sqrt{3}\int\!\frac{d^{4}q}{(2\pi)^{4}}\,\mathrm{Tr}\!\left[\Psi(K,q)\,\mathcal{O}(q)\right],
\label{eq:BSamp4d}
\end{equation}
the $\sqrt3$ being the color factor of the pair. Decomposing $q$ into a component along $K$ and
a transverse part $\hat q$ with $\hat q\!\cdot\!K=0$, and using that the $b\bar b$ interaction is
instantaneous to the accuracy considered here, the relative-energy dependence integrates out and
defines the equal-time Salpeter wave function
$\psi(\hat q)=(i/2\pi)\int\!dq_{\parallel}\Psi(K,q)$~\cite{Bhatnagar:2013bha,Gebrehana:2019mpw}.
The amplitude then reduces to the three-dimensional convolution
\begin{equation}
A=-i\sqrt{3}\int\!\frac{d^{3}\hat q}{(2\pi)^{3}}\,
\mathrm{Tr}\!\left[\psi(\hat q)\,\mathcal{O}(\hat q)\right],
\qquad \psi(\hat q)=\chi(\hat q)\,\mathcal{P}(\hat q),
\label{eq:BSamp}
\end{equation}
in which the scalar radial function $\chi(\hat q)$ carries the entire normalization and the Dirac
projector $\mathcal{P}(\hat q)$ the spin structure. Ordering the $1^{--}$ covariants in powers of
$1/M$~\cite{Bhatnagar:2013bha} and retaining those through first order in $\hat q$,
\begin{equation}
\mathcal{P}(\hat q)=M\slashed{\eps}-\slashed{\eps}\slashed{K}
+\frac{M}{m_{b}}(\hat q\!\cdot\!\eps)-\frac{1}{m_{b}}\slashed{K}(\hat q\!\cdot\!\eps)
+\frac{1}{m_{b}}\slashed{K}\slashed{\eps}\,\slashed{\hat q},
\label{eq:proj}
\end{equation}
with $\eps\equiv\eps(K)$ the polarization vector of the $\Upsilon_{n}$. The first two terms give
the nonrelativistic $^{3}S_{1}$ component and those linear in $\hat q$ are induced by the
relative motion of the pair.

Since $\chi(\hat q)$ is sharply peaked at the origin, the trace is expanded to second order,
\begin{equation}
\mathrm{Tr}[\mathcal{P}(\hat q)\mathcal{O}(\hat q)]=\mathcal{T}_{0}
+\hat q^{\mu}\mathcal{T}_{\mu}+\hat q^{\mu}\hat q^{\nu}\mathcal{T}_{\mu\nu}+O(\hat q^{3}),
\label{eq:taylor}
\end{equation}
where the linear term vanishes by parity and, $\chi(\hat q)$ being isotropic, the quadratic term
survives only through its directional average,
\begin{equation}
\langle\hat q^{\mu}\hat q^{\nu}\rangle_{\rm dir}=|\bm{\hat q}|^{2}\,P_{T}^{\mu\nu},\qquad
P_{T}^{\mu\nu}=\frac{1}{3}\Big(-g^{\mu\nu}+\frac{K^{\mu}K^{\nu}}{M^{2}}\Big).
\label{eq:PT}
\end{equation}
Carrying out the $\hat q$ integration, the bound-state dependence is expressed through the
wave function and its Laplacian at the origin,
\begin{equation}
\int\!\frac{d^{3}\hat q}{(2\pi)^{3}}\,\chi(\hat q)=\sqrt{\frac{1}{16\pi M}}\,R_{\Upsilon_{n}}(0),
\qquad
\int\!\frac{d^{3}\hat q}{(2\pi)^{3}}\,|\bm{\hat q}|^{2}\chi(\hat q)
=-\sqrt{\frac{1}{16\pi M}}\,\nabla^{2}R_{\Upsilon_{n}}(0),
\label{eq:moments}
\end{equation}
so that the amplitude takes the compact form
\begin{equation}
A=-i\sqrt{3}\,\sqrt{\frac{1}{16\pi M}}\,\Big[\,R_{\Upsilon_{n}}(0)\,\mathcal{T}_{0}
-\nabla^{2}R_{\Upsilon_{n}}(0)\,P_{T}^{\mu\nu}\mathcal{T}_{\mu\nu}\,\Big].
\label{eq:Aintegrated}
\end{equation}
The first term is the leading order and the second the order-$q^{2}$ relativistic correction,
the object of the present work, while the neglected terms are set by the higher moments of the
wave function. The effect of the correction on the rate depends on the ratio $\nabla^{2}R(0)/R(0)$ together
with the hard coefficients and its interference with the leading amplitude.
The opposite signs of this ratio for the $\Upsone$ and $\Upstwo$
are discussed in Sec.~\ref{subsec:inputs}.

The hard kernel is the amplitude for $b\bar b\rightarrow\gamma g^{*}g^{*}$, with the photon and the
two gluons attached to the bottom line in all six orderings,
\begin{equation}
\mathcal{O}(\hat q)=iQ_{b}e\,g_{s}^{2}\,\frac{\delta^{ab}}{6}\,
\slashed{\eps}^{*}(k_{2})\,S_{F}(l_{2})\,\slashed{\eps}^{*}(k_{1})\,S_{F}(l_{1})\,
\slashed{\eps}^{*}(k)+(\text{5 permutations}),
\label{eq:hardkernel}
\end{equation}
where $\eps^{*}(k)$ denotes the photon polarization vector and $\eps^{*}(k_{1,2})$ denote
the polarization vectors associated with the two virtual gluons. The gluon momenta and
color indices are $k_{1,2}$ and $a,b$, respectively, and
$S_{F}(l)=(\slashed{l}+m_{b})/(l^{2}-m_{b}^{2})$ denotes the bottom-quark propagator,
with $l_{1}=K/2+\hat q-k$ and $l_{2}=K/2+\hat q-k-k_{1}$.

The light $q\bar q$ pair is projected onto the meson by the twist-2 light-cone matrix
element~\cite{Chernyak:1983ej}
\begin{equation}
\langle\etapp(p)|\bar q_{\alpha}(z)q_{\beta}(y)|0\rangle
=\frac{i}{4}f_{\etapp}^{q}\,(\slashed{p}\gamma_{5})_{\beta\alpha}
\int_{0}^{1}\!du\,e^{i(\bar u\,p\cdot y+u\,p\cdot z)}\phi^{q}(u)+\cdots,
\label{eq:quarkMT}
\end{equation}
with $f^{q}_{\etapp}$ the flavor decay constant, $\bar u=1-u$, higher twists omitted, and the
twist-2 distribution amplitude (DA)
\begin{equation}
\phi^{q}(u)=6u\bar u\Big[1+\!\!\sum_{n=2,4,\dots}\!\!c^{q}_{n}(\mu)\,C^{3/2}_{n}(2u-1)\Big],
\label{eq:quarkDA}
\end{equation}
whose Gegenbauer coefficients $c^{q}_{n}(\mu)$ mix under evolution with those of the two-gluon DA
introduced below~\cite{Kroll:2002nt}. In the quark-flavor basis the $\eta$ and $\etap$ constants
follow from $f_{q}$, $f_{s}$, and the mixing angle $\phi$~\cite{Feldmann:1998vh}. Convolving the hard
$g^{*}g^{*}\rightarrow q\bar q$ subprocess with the DA gives the effective vertex
\begin{equation}
M^{\mu\nu}=-i\,(4\pi\alpha_{s})\,\delta_{ab}\,\epsilon^{\mu\nu\rho\sigma}k_{1\rho}k_{2\sigma}
\sum_{q}\frac{f^{q}_{\etapp}}{6}\int_{0}^{1}\!du\,
\left(\frac{\phi^{q}(u)}{\bar u k_{1}^{2}+u k_{2}^{2}-u\bar u m^{2}-m_{q}^{2}}
+(u\leftrightarrow\bar u)\right).
\label{eq:ggeta}
\end{equation}
The two subamplitudes are joined through the gluon propagators and the loop integration,
\begin{equation}
M_{T}=\frac{1}{2}\int\!\frac{d^{4}k_{1}}{(2\pi)^{4}}\,A^{\mu\nu}M_{\mu\nu}\,
\frac{i}{k_{1}^{2}}\,\frac{i}{k_{2}^{2}},
\label{eq:assemble}
\end{equation}
with $k_{2}=p-k_{1}$, $A^{\mu\nu}$ the amplitude~(\ref{eq:Aintegrated}) stripped of the gluon
polarization vectors, and the factor $1/2$ compensating the interchange of the identical gluons.
Since $\Upsilon_{n}(1^{--})\rightarrow\gamma(1^{-})+\etapp(0^{-+})$ admits the single parity-odd
structure $\epsilon^{\alpha\beta\mu\nu}K_{\mu}k_{\nu}$, one helicity amplitude survives, extracted
by the projector~\cite{Korner:1982vg}
\begin{equation}
h^{\alpha\beta}=-\frac{i}{2\,(K\!\cdot\!k)}\,\epsilon^{\alpha\beta\mu\nu}K_{\mu}k_{\nu},
\qquad H^{q}=h^{\alpha\beta}T_{\alpha\beta},
\label{eq:helproj}
\end{equation}
$T^{\alpha\beta}$ denoting $M_{T}$ with the $\Upsilon_{n}$ and photon polarizations removed.
Collecting the traces, the flavor sum and the loop integration gives
\begin{equation}
H^{q}=\frac{2Q_{b}}{2\sqrt{3}}\sqrt{4\pi\alpha_{e}}\,(4\pi\alpha_{s})^{2}\sqrt{\frac{1}{16\pi M}}\,
\sum_{q=u,d,s}f^{q}_{\etapp}\int_{0}^{1}\!du\,\phi^{q}(u)\Big[\,R_{\Upsilon_{n}}(0)\,I^{q}_{0}(u)
-\nabla^{2}R_{\Upsilon_{n}}(0)\,I^{q}_{2}(u)\,\Big],
\label{eq:Hq}
\end{equation}
with $Q_{b}=-1/3$ the bottom-quark charge. This expression has the same form as its
charmonium counterpart~\cite{He:2026yex}, with $Q_{c}\rightarrow Q_{b}$,
$R_{\psi_{n}}\rightarrow R_{\Upsilon_{n}}$, and the corresponding bottomonium mass $M$.

The leading-order kernel is the one-loop integral over five propagators,
\begin{equation}
I^{q}_{0}(u)=\int\!\frac{d^{4}k_{1}}{(2\pi)^{4}}\,\frac{1}{G_{1}G_{2}L}
\left[\frac{N_{A}}{P_{0}P_{1}}+\frac{N_{B}}{P_{0}P_{2}}+\frac{N_{C}}{P_{1}P_{2}}\right]
+(u\leftrightarrow\bar u),
\label{eq:integq0}
\end{equation}
where $G_{1}=k_{1}^{2}$ and $G_{2}=k_{2}^{2}$ are the gluon propagator denominators,
$L=(k_{1}-up)^{2}-m_{q}^{2}$ is the light-quark propagator denominator at momentum fraction $u$, and
\begin{equation}
P_{0}=(k-K/2)^{2}-m_{b}^{2},\qquad
P_{1}=(k_{1}-K/2)^{2}-m_{b}^{2},\qquad
P_{2}=(k_{1}+k-K/2)^{2}-m_{b}^{2}
\label{eq:heavydenom}
\end{equation}
are the bottom-quark propagator denominators. The $+i\epsilon$ prescription is understood for
all propagator denominators. The six diagrams collapse pairwise into the three
topologies $X\in\{A,B,C\}$ labeled by the position of the photon vertex, two boxes in which the
loop-independent $P_{0}$ factors out and one pentagon in which both $k_{1}$-dependent bottom-quark propagators
stay off shell. In the weak-binding limit $m_{b}=M/2$ the numerators reduce to polynomials in the
loop invariants, of which the box one reads
\begin{equation}
N_{A}=32i\,M\Big[(k\!\cdot\!k_{1})^{2}-(k\!\cdot\!k_{1})(k_{1}\!\cdot\!K)
+k_{1}^{2}\,k\!\cdot\!K\Big],
\label{eq:NA}
\end{equation}
with $N_{B}=N_{A}(k_{1}\rightarrow p-k_{1})$ by Bose symmetry and $N_{C}$ mapped onto itself. The
remaining forms, identical to the charmonium ones since they involve only $M$ and $k\!\cdot\!K$,
are given in Ref.~\cite{He:2026yex}. Individual topologies are not gauge invariant, only
their sum being physical, and that sum reproduces the compact loop function of
Ref.~\cite{He:2019mpy}.

$I^{q}_{0}$ is complex, its absorptive part generated by the unitarity cuts of
Eq.~(\ref{eq:integq0}). The leading one places the two gluon lines $G_{1}$ and $G_{2}$ on shell,
so that the $\etapp$ is formed from a physical gluon pair. Every cut of the integral sits in a
channel whose invariant is proportional to the meson mass, the two-gluon cut in the channel of
invariant mass squared $m^{2}$ and the quark--gluon cuts of the $g^{*}g^{*}\rightarrow\etapp$
subgraph in channels of invariant $u^{2}m^{2}$ or $\bar u^{2}m^{2}$, while the bottom pair rests
exactly at its threshold $4m_{b}^{2}=M^{2}$ and opens no phase space at all. The absorptive part
therefore closes off as $x\rightarrow0$.

Its approach to that limit is not a pure power, because the two-gluon cut reaches into a region
in which one gluon is soft. On the cut the bottom line adjacent to the $\Upsilon_{n}$ vertex
reduces to $P_{1}=-zM^{2}/2$, where $z=2k_{1}\!\cdot\!K/M^{2}$ is the energy fraction of that
gluon, and the cut kinematics confine $z$ to $x\le z\le1$, the lower limit being set by the meson
mass itself. Carrying a single power of $P_{1}$, the leading-order integrand picks up
$\int_{x}^{1}dz/z=\ln(1/x)$ from this region. Unitarity thus forces the absorptive part to vanish
and the soft-gluon region supplies the logarithm, the vanishing being linear in $x$ up to that
factor, $\mathrm{Im}\,I^{q}_{0}\propto x\ln(1/x)$, as the amplitudes of
Sec.~\ref{subsec:amplitudes} bear out. The leading-order amplitude accordingly becomes
progressively real as $x$ decreases.

The $q^{2}$ kernel follows from the transverse projection of the second
$\hat q$ derivative, Eq.~(\ref{eq:PT}). The derivative acts on the bottom
propagators and the $\hat q$-dependent numerator, including the bound-state
projector, while $G_1$, $G_2$, and $L$ remain unchanged. Organizing the result by the powers
$(a,b)$ of the bottom denominators, with $D^{A}_{a,b}=P_{0}^{a}P_{1}^{b}$,
$D^{B}_{a,b}=P_{0}^{a}P_{2}^{b}$ and $D^{C}_{a,b}=P_{1}^{a}P_{2}^{b}$ running over $a,b\ge1$ with
$a+b\le4$,
\begin{equation}
I^{q}_{2}(u)=\int\!\frac{d^{4}k_{1}}{(2\pi)^{4}}\,\frac{1}{G_{1}G_{2}L}
\sum_{X\in\{A,B,C\}}\sum_{(a,b)}\frac{N^{X}_{a,b}(k_{1})}{D^{X}_{a,b}}+(u\leftrightarrow\bar u).
\label{eq:integq2}
\end{equation}
The bottom-line denominator structure of $I^{q}_{2}$ is thus considerably richer than that of
$I^{q}_{0}$, a fact that will matter for the endpoint behavior discussed below. The eighteen
numerators $N^{X}_{a,b}$ are listed in Ref.~\cite{He:2026yex}. The pentagon propagators being
linearly dependent, the integrals are reduced by partial fractioning over $k_{1}$ with
\textsc{FeynCalc}~\cite{Mertig:1990an,Shtabovenko:2016sxi,Shtabovenko:2020gxv} and evaluated with
\textsc{Package-X}~\cite{Patel:2015tea,Patel:2016fam}.

As in the charmonium analysis~\cite{He:2019mpy,He:2026yex}, the quark loop
kernels are relatively flat in the central momentum-fraction region, with
more pronounced variation toward the endpoints. Together with the endpoint
suppression of the DA, this behavior explains the weak dependence of the
helicity amplitudes on the DA shapes considered here.

As $u\rightarrow0$ the light-quark propagator degenerates
with the gluon one, $L\rightarrow k_{1}^{2}-m_{q}^{2}$ against $G_{1}=k_{1}^{2}$, and the
integrand is enhanced in the region of small $L$. The endpoint enhancement is more pronounced
in $I^q_2$ than in $I^q_0$, but its effect on the convolution is limited by the vanishing
of the leading-twist DA at the endpoints. For the asymptotic DA $\phi^q(u)=6u\bar u$,
the combined contribution from the endpoint regions $u<0.05$ and $u>0.95$ is only
a few percent of the full convolution in magnitude.
The numerical tabulation must resolve this endpoint enhancement. The
insensitivity of the amplitude to the DA shape and to $m_{q}$ therefore carries over from
charmonium to bottomonium, as the results of Sec.~\ref{sec:numerical} confirm.

\subsection{Gluon-content contribution}
\label{subsec:gluon}

\begin{figure}[H]
\centering
\includegraphics[width=0.5\textwidth]{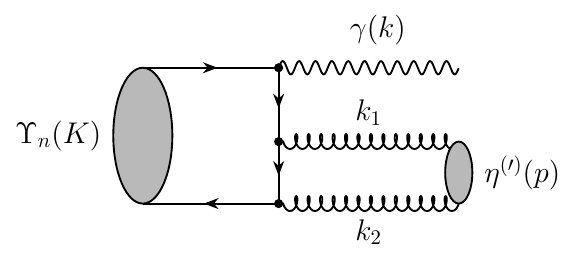}
\caption{A representative Feynman diagram for the gluon-content contribution to
$\Upsilon_{n}\rightarrow\gamma\etapp$, in which the $b\bar b$ pair annihilates into a photon
and two collinear gluons that directly form the two-gluon content of the $\etapp$. Two further
diagrams are obtained by permuting the photon and gluon legs.}
\label{fig:gluon}
\end{figure}

The $\etapp$ also carries an intrinsic two-gluon Fock component. Here the two gluons of the hard
subprocess couple to it directly, without the intermediate light-quark loop, through the
leading-twist matrix element~\cite{Kroll:2002nt,Ball:2007hb,Agaev:2014wna}
\begin{equation}
\langle\etapp(p)|A^{a}_{\alpha}(z)A^{b}_{\beta}(y)|0\rangle
=\frac{1}{4}\epsilon_{\alpha\beta\rho\sigma}\frac{k^{\rho}p^{\sigma}}{p\cdot k}
\frac{C_{F}}{\sqrt3}\frac{\delta^{ab}}{8}\,f^{g}_{\etapp}
\int_{0}^{1}\!du\,e^{i(u\,p\cdot z+\bar u\,p\cdot y)}\frac{\phi^{g}(u)}{u\bar u},
\label{eq:gluonMT}
\end{equation}
with $f^{g}_{\etapp}=\tfrac{1}{\sqrt3}(f^{u}_{\etapp}+f^{d}_{\etapp}+f^{s}_{\etapp})$ and the
gluonic twist-2 DA
\begin{equation}
\phi^{g}(u)=30\,u^{2}\bar u^{2}\!\!\sum_{n=2,4,\dots}\!\!c^{g}_{n}(\mu)\,C^{5/2}_{n-1}(2u-1).
\label{eq:gluonDA}
\end{equation}
The gluons now carry the fixed collinear momenta $up$ and $\bar u p$, so this contribution is
tree level, with a representative diagram shown in Fig.~\ref{fig:gluon}. Projecting as before,
\begin{equation}
H^{g}=\frac{2Q_{b}}{9}\sqrt{4\pi\alpha_{e}}\,(4\pi\alpha_{s})\,\frac{1}{\sqrt{\pi M}}
\frac{f^{g}_{\etapp}}{M}\int_{0}^{1}\!du\,\frac{\phi^{g}(u)}{u\bar u}\Big[\,R_{\Upsilon_{n}}(0)\,
I^{g}_{0}(u)-\nabla^{2}R_{\Upsilon_{n}}(0)\,I^{g}_{2}(u)\,\Big].
\label{eq:Hg}
\end{equation}
At leading order only the two box topologies survive,
\begin{equation}
I^{g}_{0}(u)=\frac{N^{A}_{g}}{P_{0}P_{u}}+\frac{N^{B}_{g}}{P_{0}P_{\bar u}},
\qquad N^{A}_{g}=\tfrac{1}{2}(k\!\cdot\!K)M^{2}u,\quad N^{B}_{g}=-N^{A}_{g}(u\leftrightarrow\bar u),
\label{eq:integg0}
\end{equation}
the pentagon vanishing because the collinear projection leaves the gluon momenta without
transverse components. The bottom denominators are now algebraic,
\begin{equation}
P_{0}=-k\!\cdot\!K,\qquad
P_{u}=-u\big[M^{2}\bar u+(k\!\cdot\!K)(u-\bar u)\big],\qquad
P_{\bar u}=-\bar u\big[M^{2}u+(k\!\cdot\!K)(\bar u-u)\big],
\label{eq:gluondenom}
\end{equation}
and none of them vanishes in the interior of the integration range, so the gluon amplitude carries
no absorptive part. Summing Eq.~(\ref{eq:integg0}) gives the compact form~\cite{He:2019mpy}
\begin{equation}
I^{g}_{0}(u)=\frac{2x(2u-1)}{1-x^{2}(1-2u)^{2}},
\label{eq:gluonI0compact}
\end{equation}
while the $q^{2}$ correction, obtained from the same $\hat q$ derivative, involves only higher
powers of the same three factors,
\begin{equation}
I^{g}_{2}(u)=\sum_{X\in\{A,B,C\}}\sum_{(a,b)}\frac{N^{X}_{g,a,b}}{D^{X}_{a,b}},
\label{eq:integg2}
\end{equation}
with the numerators collected in Ref.~\cite{He:2026yex}. The relative momentum revives the
pentagon, though only through its lowest term, and no new threshold opens.

Equation~(\ref{eq:gluonI0compact}) exhibits the two features that govern this contribution. The
kernel is odd under $u\leftrightarrow\bar u$, matching the antisymmetry of $\phi^{g}$, and carries
an overall factor $x$, a reflection of the near-on-shell gluons and of the Ore--Powell form of
their coupling to the $\etapp$~\cite{Krammer:1978qp,Billoire:1978xt}. The factor $x$, smaller here by an order of magnitude, strongly suppresses
the gluon amplitude and limits the sensitivity of these rates to the
gluonic component of the light mesons. Because the kernel is odd rather than flat, the convolution weights the shape of
$\phi^{g}$ instead of its normalization, which makes the gluon term the only one of the three that
responds appreciably to the DA model. Carrying no light-quark line, it is however strictly
independent of $m_{q}$.

\subsection{QED contribution}
\label{subsec:qed}

\begin{figure}[H]
\centering
\includegraphics[width=0.9\textwidth]{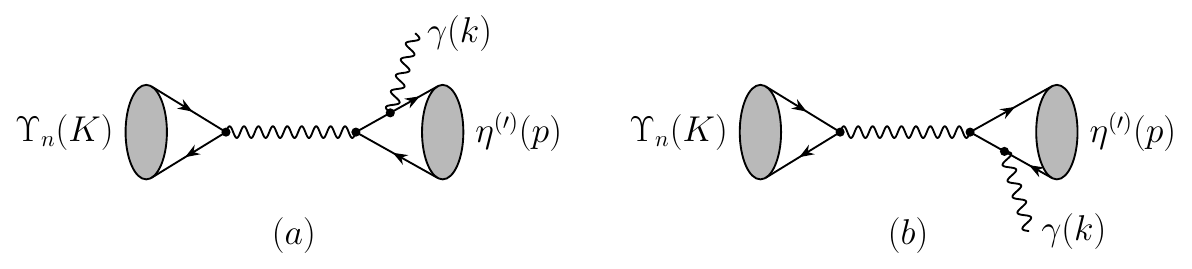}
\caption{Feynman diagrams for the QED contribution
$\Upsilon_{n}\rightarrow\gamma^{*}\rightarrow\gamma\etapp$, in which the $b\bar b$ pair
annihilates into a single virtual photon that produces the light $q\bar q$ pair of the
$\etapp$. In contrast to the quark-antiquark and two-gluon contributions, the real photon is
radiated from a final-state light-quark line, the quark in (a) and the antiquark in (b).}
\label{fig:qed}
\end{figure}

Finally, the $\etapp$ can be produced electromagnetically. The $b\bar b$ pair annihilates into a
single virtual photon, $\Upsilon_{n}\rightarrow\gamma^{*}$, which creates the light $q\bar q$ pair,
and the real photon is radiated from that light-quark line, leaving one internal light-quark
propagator, and the pair is projected onto the meson by the same matrix
element~(\ref{eq:quarkMT}). The process is tree level and is shown in Fig.~\ref{fig:qed}. It
shares the helicity structure and the bound-state projection of the quark contribution, with the
two-gluon exchange replaced by a single virtual photon, so that it is of order $\alpha_{e}$
instead of $\alpha_{s}^{2}$ and carries the squared light-quark charges $Q_{q}^{2}$. Because the
$b\bar b$ pair annihilates at a single vertex, the bottom line carries no internal propagator, the
hard amplitude is independent of $\hat q$, and the $q^{2}$ correction vanishes identically,
leaving the QED contribution purely at leading order,
\begin{equation}
H^{\rm QED}=-Q_{b}\,(4\pi\alpha_{e})^{3/2}\sqrt{\frac{3}{\pi M}}\,R_{\Upsilon_{n}}(0)
\sum_{q=u,d,s}\frac{Q_{q}^{2}f^{q}_{\etapp}}{M}\int_{0}^{1}\!du\,\phi^{q}(u)\,I^{\rm QED}(u),
\label{eq:Hqed}
\end{equation}
with the $\hat q$-independent kernel
\begin{equation}
I^{\rm QED}(u)=\frac{1-x}{2}
\left(\frac{1}{u-u\bar u\,x-m_{q}^{2}/M^{2}+i\epsilon}+(u\leftrightarrow\bar u)\right).
\label{eq:IQED}
\end{equation}
As in the charmonium case~\cite{He:2026yex} the kernel is even under
$u\leftrightarrow\bar u$ and its absorptive part, arising near $u\simeq m_{q}^{2}/M^{2}$, is
tiny and endpoint-suppressed, so the QED amplitude too is set mainly by the decay constants.
Although each contribution carries one power of the heavy-quark charge, the relative weight of
the three differs markedly from charmonium. The quark and gluon amplitudes scale as
$Q_{b}\,\alpha_{s}^{2}$ and $Q_{b}\,\alpha_{s}\,x$, both suppressed at the bottom
scale by the smaller coupling and, for the gluon term, by the small $x$, whereas the QED
amplitude scales as $Q_{b}\,\alpha_{e}\sum_{q}Q_{q}^{2}f^{q}$ and is not. The QED contribution
is therefore relatively enhanced in the bottomonium decays, as quantified in
Sec.~\ref{sec:numerical}.

%==================================================================
\section{Numerical analysis}
\label{sec:numerical}

\subsection{Decay width and input parameters}
\label{subsec:inputs}

The three contributions of Sec.~\ref{sec:framework} add coherently to a single helicity
amplitude, $H=H^{q}+H^{g}+H^{\rm QED}$, and the partial width is its modulus squared weighted by
the two-body phase space,
\begin{equation}
\Gamma(\Upsilon_{n}\rightarrow\gamma\etapp)=\frac{M^{2}-m^{2}}{16\pi M^{3}}\,\frac{2}{3}\,|H|^{2},
\label{eq:width}
\end{equation}
the factor $2/3$ being the sum over the two nonvanishing helicity configurations divided by the
three polarizations of the initial $\Upsilon_{n}$.

To organize the relativistic correction we write $H=H_{\rm LO}+H_{q^{2}}$, with $H_{\rm LO}$
governed by $R_{\Upsilon_{n}}(0)$ and $H_{q^{2}}$ by $\nabla^{2}R_{\Upsilon_{n}}(0)$ as in
Eq.~(\ref{eq:Aintegrated}). Accordingly $\Gamma_{\rm LO}$ denotes the width built from the
leading-order amplitude alone, and $\Gamma_{q^{2}}$ the width including the correction in the
fixed-order convention, that is, retaining the interference between $H_{\rm LO}$ and
$H_{q^{2}}$ but dropping the $|H_{q^{2}}|^{2}$ term, which is of relative order $v^{4}$ and lies
beyond the accuracy of the present calculation,
\begin{equation}
\Gamma_{\rm LO}=\frac{M^{2}-m^{2}}{16\pi M^{3}}\,\frac{2}{3}\,|H_{\rm LO}|^{2},\qquad
\Gamma_{q^{2}}=\frac{M^{2}-m^{2}}{16\pi M^{3}}\,\frac{2}{3}\,
\big(|H_{\rm LO}|^{2}+2\,\mathrm{Re}\big[H_{\rm LO}^{*}H_{q^{2}}\big]\big).
\label{eq:fixedorder}
\end{equation}
The masses $M_{\Upsone}=9.4604$~GeV and $M_{\Upstwo}=10.0234$~GeV, the measured total widths
$\Gamma_{\Upsone}=54.02(1.25)$~keV and $\Gamma_{\Upstwo}=31.98(2.63)$~keV, the meson masses
$m_{\eta}=0.5479$~GeV and $m_{\etap}=0.9578$~GeV, the fine-structure constant $\alpha_{e}=1/137$,
and the strong coupling $\alpha_{s}(M_{Z})=0.1180$ are taken from the
Particle Data Group~\cite{ParticleDataGroup:2024cfk}. The coupling is run at one loop with five active flavors to
the hard scale $\mu=M/2$ of each decay, which gives $\alpha_{s}(M_{\Upsone}/2)=0.206$ and
$\alpha_{s}(M_{\Upstwo}/2)=0.203$.

The leading-order amplitude is governed by the wave function at the origin and the relativistic
correction by its Laplacian. Both are fixed from the frozen-$\alpha_{s}$ Cornell potential of
Eichten and Quigg, a Coulomb-plus-linear form whose coupling runs at short distance and
saturates at long distance~\cite{Eichten:2019gig,Eichten:1978tg}. The values at the origin
$|R_{\Upsone}(0)|^{2}=5.8588~\mathrm{GeV}^{3}$ and
$|R_{\Upstwo}(0)|^{2}=2.8974~\mathrm{GeV}^{3}$ are taken from the tabulation of
Ref.~\cite{Eichten:2019hbb}. We estimate the ratio
$\nabla^{2}R_{\Upsilon_n}(0)/R_{\Upsilon_n}(0)$
from the energy eigenvalues of the same potential through the Gremm--Kapustin
relation~\cite{Gremm:1997dq}, obtaining
\begin{equation}
\frac{\nabla^{2}R_{\Upsone}(0)}{R_{\Upsone}(0)}=0.309~\mathrm{GeV}^{2},\qquad
\frac{\nabla^{2}R_{\Upstwo}(0)}{R_{\Upstwo}(0)}=-2.398~\mathrm{GeV}^{2}.
\label{eq:lapR}
\end{equation}
The opposite signs reflect the different binding of the two radial states in the adopted
potential. The compact $\Upsone$ is more strongly affected by the short-distance Coulomb
attraction, whereas the more extended $\Upstwo$ probes the confining part more strongly.
The resulting energy eigenvalues have opposite signs and, through the Gremm--Kapustin
relation, give opposite signs for $\nabla^{2}R(0)/R(0)$. This sign difference between the two
bottomonium states is absent for the $J/\psi$ and $\psi(2S)$, whose corresponding ratios are
both negative~\cite{He:2026yex}. The signs and relative magnitudes of the ratios for the
$\Upsone$ and $\Upstwo$ are also consistent with the leptonic-width analysis of
Ref.~\cite{Chung:2010vz}. With the similar hard-amplitude structure of the two bottomonium
states, the opposite signs lead to a suppression of the $\Upsone$ rates and an enhancement of
the $\Upstwo$ rates, as discussed in Sec.~\ref{subsec:amplitudes}.

The wave function at the origin, which controls the leading-order amplitude, also determines
the leptonic widths, whose measured values offer a further check,
\begin{equation}
\Gamma(\Upsilon_{n}\rightarrow e^{+}e^{-})=
\frac{4\alpha_{e}^{2}Q_{b}^{2}}{M^{2}}\,|R_{\Upsilon_{n}}(0)|^{2}
\left(1-\frac{16}{3\pi}\alpha_{s}\right),
\label{eq:leptonic}
\end{equation}
with the first-order quantum chromodynamics (QCD) correction of Ref.~\cite{Mackenzie:1981sf}. The leading-order
expression moderately overestimates the measured widths, while the negative one-loop QCD
correction brings both results below the data. This correction largely cancels in the ratio,
which is approximately

\begin{equation}
\frac{\Gamma(\Upstwo\rightarrow e^{+}e^{-})}{\Gamma(\Upsone\rightarrow e^{+}e^{-})}
\simeq\frac{|R_{\Upstwo}(0)|^{2}}{|R_{\Upsone}(0)|^{2}}\,
\frac{M_{\Upsone}^{2}}{M_{\Upstwo}^{2}}=0.44,
\label{eq:leptonicratio}
\end{equation}
close to the measured $0.457(10)$~\cite{ParticleDataGroup:2024cfk}. This agreement supports
the relative size of the two wave functions at the origin, although the individual leptonic
widths remain sensitive to perturbative corrections.

For the light mesons, we adopt the three DA models used in the charmonium
analysis~\cite{He:2019mpy,He:2026yex}, with their Gegenbauer moments at $\mu_{0}=1$~GeV
listed in Table~\ref{tab:lcda}. These sets were constructed in Ref.~\cite{Agaev:2014wna}
with quark moments spanning the range of commonly used pion DA models and gluon moments
fitted to the $\eta$ and $\etap$ transition form factors. They describe the large-$Q^{2}$ data
comparably well while allowing different DA shapes, providing a phenomenologically motivated
basis for testing the sensitivity of our predictions. We evolve the moments to $\mu=M/2$
using the Efremov--Radyushkin--Brodsky--Lepage evolution equations, including quark--gluon mixing~\cite{Kroll:2002nt}.

\begin{table}[H]
\centering
\caption{The three DA models~\cite{He:2019mpy}, specified by the Gegenbauer moments $c^{q}_{2}$,
$c^{q}_{4}$ of the quark DA and $c^{g}_{2}$ of the two-gluon DA at $\mu_{0}=1$~GeV.}
\label{tab:lcda}
\begin{ruledtabular}
\begin{tabular}{lccc}
Model & $c^{q}_{2}$ & $c^{q}_{4}$ & $c^{g}_{2}$ \\
\hline
I   & $0.10$ & $\phantom{-}0.10$  & $-0.26$ \\
II  & $0.20$ & $\phantom{-}0.00$  & $-0.31$ \\
III & $0.25$ & $-0.10$ & $-0.25$ \\
\end{tabular}
\end{ruledtabular}
\end{table}

The flavor decay constants $f^{q}_{\etapp}$ follow from the $\eta$--$\etap$ mixing. In the
quark-flavor basis the physical states are a rotation of $\eta_{q}=(u\bar u+d\bar d)/\sqrt2$ and
$\eta_{s}=s\bar s$ by a single angle $\phi$, and in the Feldmann--Kroll--Stech scheme~\cite{Feldmann:1998vh}
introduced in Sec.~\ref{subsec:quark} the flavor constants are set by $f_{q}$, $f_{s}$, and
$\phi$,
\begin{equation}
f^{u,d}_{\eta}=\frac{f_{q}}{\sqrt2}\cos\phi,\quad f^{s}_{\eta}=-f_{s}\sin\phi,\qquad
f^{u,d}_{\etap}=\frac{f_{q}}{\sqrt2}\sin\phi,\quad f^{s}_{\etap}=f_{s}\cos\phi.
\label{eq:fks}
\end{equation}
These are phenomenological quantities that vary between extractions, chiefly in the angle, so we
carry two representative determinations through the calculation, listed in
Table~\ref{tab:mixing}. The
lattice values~\cite{Ottnad:2025zxq} have the larger angle $\phi\simeq39^{\circ}$ and the
transition-form-factor values~\cite{Escribano:2013kba} the smaller $\phi\simeq33.5^{\circ}$, and
we take the lattice scheme as default. The near cancellation in the flavor combination of
the $\gamma\eta$ amplitude makes the ratio $\mathcal{R}_{nS}$ sensitive to the mixing angle,
as discussed in Sec.~\ref{subsec:BR}.

\begin{table}[H]
\centering
\caption{The two $\eta$--$\etap$ mixing determinations in the quark-flavor
basis~\cite{Feldmann:1998vh}, the lattice determination of Ref.~\cite{Ottnad:2025zxq} and the
transition-form-factor determination of Ref.~\cite{Escribano:2013kba}, denoted by $\etap$TFF.}
\label{tab:mixing}
\begin{ruledtabular}
\begin{tabular}{lccc}
Scheme & $f_{q}$~[GeV] & $f_{s}$~[GeV] & $\phi$~[$^{\circ}$] \\
\hline
Lattice     & $0.1386(44)$ & $0.1707(33)$ & $39.3(2.0)$ \\
$\etap$TFF  & $0.1419(26)$ & $0.1250(52)$ & $33.5(9)$  \\
\end{tabular}
\end{ruledtabular}
\end{table}

\subsection{Helicity amplitudes}
\label{subsec:amplitudes}

We present the helicity amplitudes in the lattice mixing scheme. For the quark and gluon
contributions, we write $H^X=H^X_{\rm LO}+H^X_{q^2}$ with $X=q,g$, while the QED contribution
has no $q^2$ term in the present approximation. Tables~\ref{tab:ampq1S}--\ref{tab:ampqed1S} give
the $\Upsone$ amplitudes for the three DA models and $m_q=0.01,0.05,0.10$~GeV.
The quark and QED amplitudes are quoted in units of $10^{-4}$~GeV and the gluon amplitudes
in units of $10^{-6}$~GeV. The gluon contribution is independent of $m_q$.

\begin{table}[H]
\centering
\caption{Quark-contribution helicity amplitude of $\Upsone\rightarrow\gamma\etapp$ in the
lattice scheme, in units of $10^{-4}$~GeV, for the three DA models and
$m_{q}=0.01,0.05,0.10$~GeV. $H^{q}_{\rm LO}$ is the leading order and $H^{q}_{q^{2}}$ the $q^{2}$
correction.}
\label{tab:ampq1S}
\begin{ruledtabular}
\begin{tabular}{llcccc}
 & & \multicolumn{2}{c}{$\gamma\eta$} & \multicolumn{2}{c}{$\gamma\etap$} \\
\cline{3-4}\cline{5-6}
Model & $m_{q}$ [GeV] & $H^{q}_{\rm LO}$ & $H^{q}_{q^{2}}$ & $H^{q}_{\rm LO}$ & $H^{q}_{q^{2}}$ \\
\hline
I   & 0.01 & $1.040+0.061i$ & $-0.024+0.025i$ & $6.264+0.856i$ & $-0.174+0.060i$ \\
I   & 0.05 & $1.040+0.059i$ & $-0.022+0.026i$ & $6.266+0.849i$ & $-0.173+0.063i$ \\
I   & 0.10 & $1.037+0.055i$ & $-0.019+0.026i$ & $6.266+0.830i$ & $-0.169+0.067i$ \\
II  & 0.01 & $1.044+0.062i$ & $-0.024+0.025i$ & $6.289+0.864i$ & $-0.176+0.061i$ \\
II  & 0.05 & $1.044+0.060i$ & $-0.022+0.026i$ & $6.290+0.858i$ & $-0.175+0.064i$ \\
II  & 0.10 & $1.041+0.055i$ & $-0.019+0.026i$ & $6.290+0.838i$ & $-0.171+0.068i$ \\
III & 0.01 & $1.046+0.062i$ & $-0.024+0.026i$ & $6.302+0.867i$ & $-0.177+0.062i$ \\
III & 0.05 & $1.046+0.060i$ & $-0.022+0.026i$ & $6.304+0.861i$ & $-0.176+0.065i$ \\
III & 0.10 & $1.044+0.056i$ & $-0.019+0.026i$ & $6.304+0.842i$ & $-0.172+0.069i$ \\
\end{tabular}
\end{ruledtabular}
\end{table}

\begin{table}[H]
\centering
\caption{Gluon-contribution helicity amplitude of $\Upsone\rightarrow\gamma\etapp$ in the
lattice scheme, in units of $10^{-6}$~GeV, for the three DA models. $H^{g}_{\rm LO}$ is the
leading order and $H^{g}_{q^{2}}$ the $q^{2}$ correction. The component carries no $m_{q}$
dependence.}
\label{tab:ampg1S}
\begin{ruledtabular}
\begin{tabular}{lcccc}
 & \multicolumn{2}{c}{$\gamma\eta$} & \multicolumn{2}{c}{$\gamma\etap$} \\
\cline{2-3}\cline{4-5}
Model & $H^{g}_{\rm LO}$ & $H^{g}_{q^{2}}$ & $H^{g}_{\rm LO}$ & $H^{g}_{q^{2}}$ \\
\hline
I   & $0.275$ & $-0.017$ & $4.943$ & $-0.307$ \\
II  & $0.325$ & $-0.020$ & $5.837$ & $-0.362$ \\
III & $0.258$ & $-0.016$ & $4.645$ & $-0.288$ \\
\end{tabular}
\end{ruledtabular}
\end{table}

\begin{table}[H]
\centering
\caption{QED-contribution helicity amplitude of $\Upsone\rightarrow\gamma\etapp$ in the lattice
scheme, in units of $10^{-4}$~GeV, for the three DA models and $m_{q}=0.01,0.05,0.10$~GeV.
$H^{\rm QED}$ is real to the precision shown and receives no $q^{2}$ correction.}
\label{tab:ampqed1S}
\begin{ruledtabular}
\begin{tabular}{llcc}
Model & $m_{q}$ [GeV] & $H^{\rm QED}_{\eta}$ & $H^{\rm QED}_{\etap}$ \\
\hline
I   & 0.01 & $0.745$ & $1.214$ \\
I   & 0.05 & $0.746$ & $1.214$ \\
I   & 0.10 & $0.747$ & $1.216$ \\
II  & 0.01 & $0.749$ & $1.219$ \\
II  & 0.05 & $0.749$ & $1.220$ \\
II  & 0.10 & $0.750$ & $1.221$ \\
III & 0.01 & $0.737$ & $1.201$ \\
III & 0.05 & $0.737$ & $1.201$ \\
III & 0.10 & $0.738$ & $1.202$ \\
\end{tabular}
\end{ruledtabular}
\end{table}

The quark contribution provides the largest individual amplitude, while the QED amplitude
reaches about $70\%$ of $|H^q_{\rm LO}|$ in $\gamma\eta$ and $20\%$ in $\gamma\etap$.
Constructive interference between them makes the electromagnetic contribution particularly
important in $\gamma\eta$. Its
relative importance reflects the smaller QCD coupling at the bottomonium scale and the different
flavor weights of the two contributions. The gluon amplitude is smaller than the quark one by
2 to 3 orders of magnitude. At the precision relevant here, only the quark contribution
has an appreciable imaginary part, generated by the on-shell intermediate states discussed in
Sec.~\ref{subsec:quark}. Its leading-order absorptive part is suppressed at small $x$, while
interference with the effectively real QED amplitude is governed by the quark amplitude's real
part.

The quark amplitude varies little across the DA models and light-quark masses in
Table~\ref{tab:ampq1S}, consistent with the weak momentum-fraction dependence of its
leading-order loop kernel. The QED amplitude in Table~\ref{tab:ampqed1S} also changes by less
than $2\%$ among the models, although its kernel weights the inverse moment of the DA. In the
small-$x$ and small-$m_q$ limit, this convolution is proportional to $1+c_2^q+c_4^q$, which is
similar for the three evolved models. By contrast, the gluon amplitude in Table~\ref{tab:ampg1S}
responds directly to the gluon Gegenbauer moment and varies by about a quarter. Its small
absolute size makes this variation unimportant for the coherent sum.

For the $\Upsone$, the real part of $H^q_{q^2}$ opposes the leading-order real part at the level
of $2\%$ to $3\%$. Its imaginary part has the same sign as the leading-order one, but their
interference makes only a small positive contribution to the width and does not offset the
decrease from the real parts. The gluon amplitude decreases by about $6\%$, with little effect
on the total rate. The net correction is therefore dominated by the quark term and lowers the
coherent widths by a few percent. This behavior follows from the small magnitude of the ratio
$\nabla^{2}R(0)/R(0)$ discussed
in Sec.~\ref{subsec:inputs}, together with the calculated hard coefficients.

Table~\ref{tab:amp2S} presents the corresponding $\Upstwo$ amplitudes for Model~I at
$m_q=0.01$~GeV. The hierarchy of the three contributions remains similar, while the relativistic
correction changes substantially.

\begin{table}[H]
\centering
\caption{Helicity amplitudes of the three contributions to $\Upstwo\rightarrow\gamma\etapp$ in
the lattice scheme, for Model~I at $m_{q}=0.01$~GeV, in units of $10^{-4}$~GeV. For each
contribution LO is the leading order and $q^{2}$ its correction, the full amplitude being their
sum.}
\label{tab:amp2S}
\begin{ruledtabular}
\begin{tabular}{lcccc}
 & \multicolumn{2}{c}{$\gamma\eta$} & \multicolumn{2}{c}{$\gamma\etap$} \\
\cline{2-3}\cline{4-5}
Contribution & LO & $q^{2}$ & LO & $q^{2}$ \\
\hline
Quark $H^{q}$     & $0.650+0.035i$ & $0.098-0.115i$ & $3.915+0.491i$ & $0.745-0.300i$ \\
Gluon $H^{g}$     & $0.0015$       & $0.0007$       & $0.0273$       & $0.0117$ \\
QED $H^{\rm QED}$ & $0.480$        & $0$            & $0.782$        & $0$ \\
\end{tabular}
\end{ruledtabular}
\end{table}

The ratio $\nabla^{2}R(0)/R(0)$ for the $\Upstwo$ is larger in magnitude and opposite in sign
to that for the $\Upsone$. Together with the similar hard-amplitude structure, this yields a
larger real correction for the $\Upstwo$ that interferes constructively with its leading-order
real part. The imaginary
correction instead opposes the leading-order imaginary part and even exceeds it in magnitude in
$\gamma\eta$, but the width remains dominated by the real-part interference. The gluon amplitude
receives a sizable relative correction of about $43\%$, yet its small absolute size again limits
its effect. The coherent widths consequently increase by about $17\%$ in $\gamma\eta$ and $30\%$
in $\gamma\etap$. The relativistic correction remains smaller in magnitude than the leading-order
amplitude, while the difference between the two radial states is clearly visible. We next examine the resulting
branching ratios and their dependence on the $\eta$--$\etap$ mixing parameters.

\subsection{Branching ratios}
\label{subsec:BR}

The branching ratios are obtained from the coherent quark, gluon, and QED amplitudes using
the fixed-order prescription of Eq.~(\ref{eq:fixedorder}). We use Model~I at
$m_q=0.01$~GeV and $\mu=M/2$. Tables~\ref{tab:BR1S} and~\ref{tab:BR2S} list
the leading-order branching ratios $\mathcal{B}_{\rm LO}$ and the values including the
$q^2$ correction, $\mathcal{B}_{q^2}$, in units of $10^{-6}$ for the two mixing
schemes, together with $\mathcal{R}_{nS}$ of Eq.~(\ref{eq:Rdef}). The quoted branching-ratio
uncertainties include the mixing-parameter and measured total-width errors. For
$\mathcal{R}_{nS}$, the total width cancels, and the mixing errors are propagated with the
two channels correlated. The common wave-function factor also cancels, although the
different powers of $\alpha_s$ in the three amplitudes leave a residual scale dependence.
Table~\ref{tab:BR1S} includes the CLEO $90\%$~C.L.\ upper limits~\cite{CLEO:2007iso}.

\begin{table}[H]
\centering
\caption{Branching ratios $\mathcal{B}(\Upsone\rightarrow\gamma\etapp)$ in units of $10^{-6}$,
for Model~I at $m_{q}=0.01$~GeV, in the lattice and $\etap$TFF mixing schemes.
$\mathcal{B}_{\rm LO}$ is the leading order and $\mathcal{B}_{q^{2}}$ the value with the $q^{2}$
correction, and the last row gives the ratio $\mathcal{R}_{1S}$. The last column is the CLEO
upper limit~\cite{CLEO:2007iso}.}
\label{tab:BR1S}
\begin{ruledtabular}
\begin{tabular}{lccccc}
 & \multicolumn{2}{c}{Lattice} & \multicolumn{2}{c}{$\etap$TFF} & \\
\cline{2-3}\cline{4-5}
 & $\mathcal{B}_{\rm LO}$ & $\mathcal{B}_{q^{2}}$ & $\mathcal{B}_{\rm LO}$ & $\mathcal{B}_{q^{2}}$ & CLEO \\
\hline
$\Upsone\rightarrow\gamma\eta$  & $0.83(28)$    & $0.81(27)$    & $2.85(28)$ & $2.76(27)$ & $<1.0$ \\
$\Upsone\rightarrow\gamma\etap$ & $14.74(70)$   & $14.08(67)$   & $10.47(54)$ & $10.00(51)$ & $<1.9$ \\
$\mathcal{R}_{1S}$              & $17.81(6.08)$ & $17.46(5.93)$ & $3.68(44)$ & $3.63(43)$ & \\
\end{tabular}
\end{ruledtabular}
\end{table}

For the $\Upsone$, the lattice result for $\gamma\eta$ is compatible with the CLEO upper
limit, whereas the $\etap$TFF central value exceeds it by about a factor of $3$. The
$\gamma\etap$ central predictions exceed the limit by factors of approximately $7$ and $5$, respectively.
The few-percent decrease from the $q^2$ correction is too small to resolve this tension.
The larger scheme dependence in $\gamma\eta$ follows from the flavor combination
$\sqrt2 f_q\cos\phi-f_s\sin\phi$ in the dominant QCD contribution, where partial
cancellation enhances sensitivity to the mixing parameters. In $\gamma\etap$, the
corresponding combination $\sqrt2 f_q\sin\phi+f_s\cos\phi$ is additive. The QED term has
different charge weights and also affects the coherent rates. These differences give
$\mathcal{R}_{1S}=17.46(5.93)$ and $3.63(43)$ in the two schemes, making a measurement of
both channels useful for distinguishing the mixing inputs within this mechanism. The
charmonium results of Ref.~\cite{He:2026yex} provide a complementary comparison, while the
present $\Upsone$ upper limits do not establish a preferred scheme.

Table~\ref{tab:BR2S} presents the corresponding $\Upstwo$ branching ratios and
$\mathcal{R}_{2S}$ with the same choices of DA model and light-quark mass.

\begin{table}[H]
\centering
\caption{Branching ratios $\mathcal{B}(\Upstwo\rightarrow\gamma\etapp)$ in units of $10^{-6}$,
for Model~I at $m_{q}=0.01$~GeV, in the lattice and $\etap$TFF mixing schemes.
$\mathcal{B}_{\rm LO}$ is the leading order and $\mathcal{B}_{q^{2}}$ the value with the $q^{2}$
correction, and the last row gives the ratio $\mathcal{R}_{2S}$.}
\label{tab:BR2S}
\begin{ruledtabular}
\begin{tabular}{lccccc}
 & \multicolumn{2}{c}{Lattice} & \multicolumn{2}{c}{$\etap$TFF} \\
\cline{2-3}\cline{4-5}
 & $\mathcal{B}_{\rm LO}$ & $\mathcal{B}_{q^{2}}$ & $\mathcal{B}_{\rm LO}$ & $\mathcal{B}_{q^{2}}$ \\
\hline
$\Upstwo\rightarrow\gamma\eta$  & $0.53(18)$   & $0.62(22)$    & $1.80(22)$ & $2.17(27)$ \\
$\Upstwo\rightarrow\gamma\etap$ & $9.25(85)$   & $12.06(1.11)$ & $6.57(62)$ & $8.56(81)$ \\
$\mathcal{R}_{2S}$              & $17.50(5.94)$ & $19.54(6.81)$ & $3.64(43)$ & $3.94(47)$ \\
\end{tabular}
\end{ruledtabular}
\end{table}

For the $\Upstwo$, the $q^2$ correction increases the rates by about $17\%$ to $30\%$, as
anticipated from the amplitude analysis in Sec.~\ref{subsec:amplitudes}. The corrected
$\gamma\etap$ branching ratios are of order $10^{-5}$, while those for $\gamma\eta$ are
of order $10^{-6}$ and again show the larger dependence on the mixing scheme. The ratios
$\mathcal{R}_{2S}=19.54(6.81)$ and $3.94(47)$ remain close to their $\Upsone$ counterparts.
This similarity reflects the common light-meson mixing inputs and the similar hierarchy
of the annihilation amplitudes. The relativistic corrections affect the two channels
unequally, producing a modest change in their ratio despite the larger changes in the
individual rates.

The preceding tables include the light-quark, gluon, and QED annihilation contributions.
The heavy-pseudoscalar admixtures of the $\etapp$ introduce additional amplitudes that can
interfere with them. We next examine these contributions and the scale dependence of the
coherent rates to assess their effect on the comparison with the $\Upsone$ limits.

\subsection{\texorpdfstring{The $\etac$-mixing and $\etab$-mixing contributions}{The eta-c-mixing and eta-b-mixing contributions}}
\label{subsec:etac}

The $U_A(1)$ anomaly can generate small $\etac$ and $\etab$ admixtures in the physical
$\eta$ and $\etap$~\cite{Chao:1989pi,Chao:1990im}. The charm component enters the one-loop
QCD and tree-level QED amplitudes through the same topologies as the light flavors, with
mass $m_c$ and weight $f^c_{\etapp}$. The bottom component considered here is reached through
the magnetic-dipole (M1) transition $\Upsilon_n\rightarrow\gamma\etab$. We denote these contributions
$H^{\etac}$ and $H^{\etab}$ and add them coherently to
$H^{\rm ann}=H^q+H^g+H^{\rm QED}$.

We estimate the admixtures through the heavy-flavor axial-current matrix element,
\begin{equation}
\langle0|\bar Q\gamma_{\mu}\gamma_{5}Q|\etapp(p)\rangle=i\,f^{Q}_{\etapp}\,p_{\mu},
\qquad Q=c,b,
\label{eq:fQme}
\end{equation}
for which we adopt the heavy-quark-loop estimate of Ref.~\cite{Ali:1997ex},
\begin{equation}
f^{Q}_{\etapp}=-\Delta i_{5}\!\left(\frac{m_{\etapp}^{2}}{m_{Q}^{2}}\right)f^{u}_{\etapp},
\label{eq:fQ}
\end{equation}
with $f^{u}_{\etapp}$ given in Eq.~(\ref{eq:fks}) and
\begin{equation}
\Delta i_{5}(z)=-1+\frac{1}{z}\left[\pi-2\arctan\sqrt{\frac{4}{z}-1}\,\right]^{2}
\;\simeq\;\frac{z}{12}\qquad(z\ll1).
\label{eq:Deltai5}
\end{equation}
The admixture is then $c^{\etapp}_{Q}=f^{Q}_{\etapp}/f_{\eta_{Q}}$. The small-$z$ limit displays
the suppression of the heavy-flavor matrix elements, $m_{\etapp}^{2}/m_{Q}^{2}$, an order of
magnitude stronger for the bottom than for the charm. Taking the current masses
$m_{c}(m_{c})=1.273(5)$~GeV and $m_{b}(m_{b})=4.183(7)$~GeV~\cite{ParticleDataGroup:2024cfk} and the
lattice decay constants $f_{\etac}=0.387(7)$~GeV~\cite{Becirevic:2013bsa} and
$f_{\etab}=0.724(12)$~GeV~\cite{Hatton:2021dvg}, the lattice mixing scheme gives
\begin{equation}
\begin{aligned}
c^{\eta}_{\etac}&=-3.10(15)\times10^{-3}, &\qquad c^{\etap}_{\etac}&=-8.20(47)\times10^{-3},\\
c^{\eta}_{\etab}&=-1.50(7)\times10^{-4},  &\qquad c^{\etap}_{\etab}&=-3.77(21)\times10^{-4}.
\end{aligned}
\label{eq:cQvalues}
\end{equation}
The corresponding charm decay constants, $f^c_\eta=-1.2$~MeV and
$f^c_{\etap}=-3.2$~MeV, agree with the heavy-quark-expansion
estimates~\cite{Franz:2000ee,Beneke:2002jn} and are much smaller than the light-flavor decay constants.

The charm contribution uses the hard amplitudes of Sec.~\ref{sec:framework}. For this
estimate we assume the same leading-twist profile as for the light flavors,
$\phi^{c}(u)\simeq\phi^{q}(u)$, while retaining $m_c$ in the hard kernel. With
$f^{c}_{\etapp}$ evaluated in each mixing scheme, the calculated $H^{\etac}$ interferes
destructively with $H^{\rm ann}$ and lowers the annihilation rates for the inputs used below
by a few percent. The small effect reflects the
$m_{\etapp}^{2}/m_{c}^{2}$ suppression. Its phase is determined by the calculation,
and no additional relative phase is introduced between $H^{\rm ann}$ and $H^{\etac}$.

The $\etab$ contribution is described by the M1 transition amplitude. Both
$\Upsilon_{n}\rightarrow\gamma\etapp$ and $\Upsilon_{n}\rightarrow\gamma\etab$ are
$1^{--}\rightarrow\gamma\,0^{-+}$ transitions carrying the single parity-odd structure of
Eq.~(\ref{eq:helproj}), so we write
\begin{equation}
\mathcal{M}^{\etab}=c^{\etapp}_{\etab}\;g_{\Upsilon_{n}\gamma\etab}\;F_{n}(k_{\etapp})\;
\epsilon_{\mu\nu\lambda\sigma}K^{\lambda}k^{\sigma}\,\eps^{\mu}(K)\,\eps^{*\nu}(k),
\label{eq:Metab}
\end{equation}
with the admixture $c^{\etapp}_{\etab}=f^{b}_{\etapp}/f_{\etab}$, the effective M1 coupling
$g_{\Upsilon_{n}\gamma\etab}$, and the radial overlap form factor $F_{n}$ evaluated at the photon
energy $k_{\etapp}=(M^{2}-m_{\etapp}^{2})/2M$ of the channel rather than at the $\etab$ point.
Applying the helicity projector gives the helicity amplitude $H^{\etab}$. The construction is
that of the charmonium analysis~\cite{He:2026yex}, to which we refer for details. The form factor
$F_{n}$ is modeled by harmonic-oscillator overlaps with their length matched to the $1S$
root-mean-square radius obtained from the frozen-$\alpha_{s}$ Cornell potential used for
$R_{\Upsilon_{n}}(0)$. The coupling is fixed by anchoring the same vertex to the on-shell M1 transition,
$|g_{\Upsilon_{n}\gamma\etab}|=|F_{n}(k_{\etab})|^{-1}
[12\pi\Gamma(\Upsilon_{n}\rightarrow\gamma\etab)/k_{\etab}^{3}]^{1/2}$, with $k_{\etab}$ the
photon energy of that transition. For the $\Upstwo$ it is the hindered transition, measured by
Belle, $\mathcal{B}=6.1(1.0)\times10^{-4}$~\cite{Belle:2018aht} using a symmetrized uncertainty,
in agreement with the lattice
nonrelativistic QCD (NRQCD) result $5.4(1.8)\times10^{-4}$~\cite{Hughes:2015dba}. We anchor to the measurement, which
together with the total width gives $\Gamma(\Upstwo\rightarrow\gamma\etab)=19.5(3.6)$~eV. For the
$\Upsone$ the transition is allowed but has not been observed, and we adopt
the screened-potential quark-model prediction
$\Gamma(\Upsone\rightarrow\gamma\etab)\simeq10$~eV of Ref.~\cite{Deng:2016ktl}.
We assign an uncertainty of $5$~eV to reflect the variation among representative
quark-model and potential nonrelativistic QCD calculations
\cite{Ebert:2002pp,Godfrey:2015dia,Segovia:2016xqb,Pineda:2013lta},
giving the input $\Gamma(\Upsone\rightarrow\gamma\etab)=10(5)$~eV used below.

The M1 width determines the magnitude of the coupling. In the charmonium analysis,
the relative phase was subsequently fitted by comparing the coherent sum with the measured
$\gamma\eta$ branching ratios~\cite{He:2026yex}. The available upper limits for the
$\Upsone$ do not support a similarly informative fit. We therefore consider fully destructive and fully constructive
interference of $H^{\etab}$ with $H^{\rm ann}+H^{\etac}$. Table~\ref{tab:etac} presents
the resulting branching ratios in both mixing schemes, illustrating the range generated
by the undetermined relative phase in each channel. The coherent branching ratios retain
the relativistic corrections to the annihilation and charm amplitudes through order $q^{2}$,
following the fixed-order prescription of Sec.~\ref{subsec:BR}. The quoted uncertainties
account for the shared inputs in the annihilation and mixing contributions.

\begin{table}[H]
\centering
\caption{Branching ratios $\mathcal{B}$ in units of $10^{-6}$, for Model~I at
$m_{q}=0.01$~GeV, from the coherent sum $H^{\rm ann}+H^{\etac}+H^{\etab}$. The two columns of
each scheme correspond to the two extreme choices of the undetermined phase of $H^{\etab}$,
destructive and constructive with respect to $H^{\rm ann}+H^{\etac}$. The last column is the
CLEO $90\%$~C.L.\ upper limit~\cite{CLEO:2007iso}. The quoted CLEO limits apply to the $\Upsone$.
An asterisk marks a negative fixed-order result caused by strong cancellation,
for which no physical branching ratio is quoted.}
\label{tab:etac}
\begin{ruledtabular}
\begin{tabular}{llccccc}
 & & \multicolumn{2}{c}{Lattice} & \multicolumn{2}{c}{$\etap$TFF} & \\
\cline{3-4}\cline{5-6}
 & & destructive & constructive & destructive & constructive & CLEO \\
\hline
$\Upsone$ & $\mathcal{B}(\gamma\eta)$   & $0.25(16)$   & $1.60(49)$    & $1.48(31)$  & $4.26(57)$   & $<1.0$ \\
          & $\mathcal{B}(\gamma\etap)$  & $7.45(1.37)$ & $21.87(2.59)$ & $5.06(1.02)$& $15.89(1.91)$& $<1.9$ \\
\hline
$\Upstwo$ & $\mathcal{B}(\gamma\eta)$   & $0.47(21)$   & $5.02(1.04)$  & $\ast$     & $9.50(1.10)$ & $\cdots$ \\
          & $\mathcal{B}(\gamma\etap)$  & $\ast$      & $50.70(6.03)$ & $\ast$     & $38.17(4.11)$& $\cdots$ \\
\end{tabular}
\end{ruledtabular}
\end{table}

Table~\ref{tab:etac} shows that the small $\etab$ admixture can nevertheless have an
appreciable effect through interference. For the $\Upsone$, the small $k_{\etab}$ enhances
the coupling extracted from the adopted M1 width, partly compensating the suppression
from the admixture and the form factor. The constructive rates are several times the
destructive ones, and the latter lower the $\gamma\etap$ rates by roughly half relative to
Table~\ref{tab:BR1S}, alleviating the excess although the central values remain above the
bound at $\mu=M/2$. For the $\Upstwo$, the coupling anchored to the measured M1 width
introduces the large ratio $F_2(k_{\etapp})/F_2(k_{\etab})$ in Eq.~(\ref{eq:Metab}),
as the $1S$--$2S$ orthogonality suppresses the overlap at the M1 point. Together with the
smaller annihilation amplitude, this makes $|H^{\etab}|$ exceed $|H^{\rm ann}|$ in the
present calculation. When destructive interference strongly suppresses the leading-order
coherent amplitude, the relativistic interference term can make the fixed-order branching ratio
negative. These negative results signal a breakdown of the truncated expansion and are marked
by asterisks rather than quoted as physical branching ratios.

The renormalization-scale dependence provides an indication of the perturbative
uncertainty at the order considered here. The quark amplitudes, including the charm QCD
term, scale as $\alpha_s^2$, and the gluon amplitude as $\alpha_s$. The QED terms and the
M1 mixing amplitude remain unchanged in this variation. Table~\ref{tab:scale} compares
$\mu=M/2$ and $\mu=M$ with destructive $\etab$ interference, rescaling these powers of
$\alpha_s$ while keeping the hard kernels and distribution-amplitude inputs fixed.

\begin{table}[H]
\centering
\caption{Branching ratios $\mathcal{B}$ in units of $10^{-6}$, for Model~I at
$m_{q}=0.01$~GeV, with the $\etac$-mixing and $\etab$-mixing contributions included and the
latter taken destructive, at the two scales $\mu=M/2$ and $\mu=M$, using the coherent
prescription of Table~\ref{tab:etac}. The last column is the CLEO
$90\%$~C.L.\ upper limit~\cite{CLEO:2007iso}. The quoted CLEO limits apply to the $\Upsone$.
An asterisk marks a negative fixed-order result caused by strong cancellation,
for which no physical branching ratio is quoted.}
\label{tab:scale}
\begin{ruledtabular}
\begin{tabular}{llccccc}
 & & \multicolumn{2}{c}{Lattice} & \multicolumn{2}{c}{$\etap$TFF} & \\
\cline{3-4}\cline{5-6}
 & & $\mu=M/2$ & $\mu=M$ & $\mu=M/2$ & $\mu=M$ & CLEO \\
\hline
$\Upsone$ & $\mathcal{B}(\gamma\eta)$  & $0.25(16)$   & $0.13(10)$ & $1.48(31)$   & $0.81(21)$ & $<1.0$ \\
          & $\mathcal{B}(\gamma\etap)$ & $7.45(1.37)$ & $3.53(93)$ & $5.06(1.02)$ & $2.36(69)$ & $<1.9$ \\
\hline
$\Upstwo$ & $\mathcal{B}(\gamma\eta)$  & $0.47(21)$ & $0.67(22)$   & $\ast$   & $0.17(13)$ & $\cdots$ \\
          & $\mathcal{B}(\gamma\etap)$ & $\ast$   & $0.88(73)$   & $\ast$   & $0.97(62)$ & $\cdots$ \\
\end{tabular}
\end{ruledtabular}
\end{table}

For the $\Upsone$, the smaller $\alpha_s$ at $\mu=M$ reduces the hard contribution and
brings the coherent branching ratios closer to the CLEO limits. The $\gamma\eta$ central values lie
below the upper limit in both schemes, while the $\gamma\etap$ result in the $\etap$TFF
scheme is compatible with the bound within the quoted uncertainty. The lattice result
also moves closer, although its central value remains above the bound. For the $\Upstwo$,
reducing the hard amplitude weakens its cancellation against the larger M1 mixing term,
making the coherent branching ratios particularly sensitive to the scale under destructive interference.

The $\etac$ contribution gives a small correction, whereas $\etab$ mixing can substantially
modify the coherent rates. Destructive interference and the scale dependence can together
alleviate the tension with the $\Upsone$ limits. Higher-order QCD calculations and measurements
of both channels would help constrain the scale sensitivity and the mixing contribution.

%==================================================================
\section{Summary}
\label{sec:summary}

Our calculation of $\Upsone,\Upstwo\rightarrow\gamma\etapp$ identifies a hierarchy in which
the quark contribution dominates, the electromagnetic contribution provides substantial
constructive interference, and the two-gluon contribution is strongly suppressed.
The smaller strong coupling at the bottomonium scale and the different flavor weights
make the electromagnetic term particularly important in $\gamma\eta$, where its
leading-order amplitude reaches about $70\%$ of the quark amplitude in the lattice scheme.
The small ratio $m_{\etapp}^{2}/M^{2}$ suppresses the gluon kernel and limits the sensitivity
of these rates to the gluonic component of the light mesons. The annihilation widths
depend only weakly on the distribution-amplitude models and light-quark masses tested,
with variations of a few percent.

Including the order-$q^{2}$ terms gives opposite changes in the two radial states.
The $\Upsone$ annihilation rates decrease by a few percent, while the $\Upstwo$ rates
increase by approximately $17\%$--$30\%$. The opposite signs of the ratios
$\nabla^{2}R(0)/R(0)$ estimated from the adopted potential, combined with the calculated hard
coefficients, produce this difference through interference with the leading-order
amplitudes. The relativistic correction to the annihilation rate remains subleading
for both states at this order, in contrast to the large effect found for the $\psip$ in
Ref.~\cite{He:2026yex}.

The weak dependence on the tested light-meson distribution amplitudes does not remove
the sensitivity to the mixing parameters. Partial cancellation between the light-flavor
components of the quark amplitude enhances this sensitivity in $\gamma\eta$ and hence
in $\mathcal{R}_{nS}$. The annihilation predictions for this ratio differ by about a
factor of $5$ between the lattice and $\etap$TFF schemes for both radial states.
The common wave-function factor cancels in the ratio, although scale dependence remains.
At $\mu=M/2$, the $\Upsone\rightarrow\gamma\eta$ central prediction lies below the
CLEO upper limit in the lattice scheme but exceeds it in the $\etap$TFF scheme.
For $\Upsone\rightarrow\gamma\etap$, the central predictions exceed the upper limit
by factors of approximately $7$ and $5$, respectively. The small relativistic
suppression cannot account for this excess,
and the comparison with the upper limits does not establish a clear preference
between the two mixing schemes.

Heavy-pseudoscalar mixing introduces additional amplitudes, while scale variation
probes the perturbative uncertainty of the annihilation contribution. The $\etac$
term gives a small destructive correction, whereas $\etab$ mixing can appreciably
modify the rates through interference. For the $\Upsone$, destructive interference
reduces the $\gamma\etap$ rates by roughly half at $\mu=M/2$. With the same destructive
choice at $\mu=M$, the $\Upsone\rightarrow\gamma\eta$ central predictions lie below
the CLEO upper limit in both schemes. The $\gamma\etap$ branching ratio in the
$\etap$TFF scheme is $2.36(69)\times10^{-6}$, compatible with its upper limit within
the quoted uncertainty. These effects therefore provide a possible way to
alleviate the discrepancy, although the coherent branching ratios depend on the M1 input and
relative phase. For the $\Upstwo$, strong cancellation can destabilize the fixed-order
expansion, limiting the reliability of the predictions for destructive interference. Higher-order QCD
calculations would help assess the scale dependence, while improved M1 inputs and
measurements of both radiative channels would help constrain the mixing contribution
and its interference.

%==================================================================
\begin{acknowledgments}
The work of J.-K.~H. is supported by the National Natural Science Foundation of China under
Grant No.~12305086, and the Project of Science and Technology Research Program of the Hubei
Provincial Department of Education under Grant No.~Q20222504. The work of C.-J.~F. is supported
by the Hubei Provincial Natural Science Foundation Youth Project under Grant No.~2024AFB151.
\end{acknowledgments}

%==================================================================
\bibliography{refs}

\end{document}